# Specular Inverse Faraday Effect in Transition Metals

Victor H. Ortiz, Shashi B. Mishra, Luat Vuong, Sinisa Coh, Richard B. Wilson[1]

Department of Mechanical Engineering and Materials Science and Engineering Program,

University of California, Riverside, California 92521, USA

## Abstract

The inverse Faraday effect is an opto-magnetic phenomenon that describes the ability of circularly polarized light to induce magnetism in solids. The capability of light to control magnetic order in solid state materials and devices is of interest for a variety of applications, such as magnetic recording, quantum computation and spintronic technologies. However, significant gaps in understanding about the effect persist, such as what material properties govern the magnitude of the effect in metals. In this work, we report time-resolved measurements of the specular inverse Faraday effect in non-magnetic metals, i.e., the magneto-optic Kerr effect induced by circularly polarized light. We measure this specular inverse Faraday effect in Cu, Pd, Pt, W, Ta, and Au at a laser wavelength of 783 nm. For Ta and W, we investigate both α and β phases. We observe that excitation of these metals with circularly polarized light induces significant circular dichroism. This nonlinear magneto-optical response to circularly polarized light is an order of magnitude larger in α-W than other metals, e.g., Pt, Au, and is greater than nearly all other reported values for the inverse Faraday effect in other materials. Our results benchmark the range of the inverse Faraday effect that can be observed in non-magnetic metals and provide insight into what material properties govern the inverse Faraday effect in metals.

[1]Corresponding author, rwilson@ucr.edu

## Introduction

The inverse Faraday effect (IFE) describes the interaction of circularly polarized light with electrons in a material. In the case of a metal, when circularly polarized light interacts with its electrons, it affects their orbital motion. Due to the coupling between spin and orbital motions through spin-orbit interaction, the spin of electrons can change [1,2]. This collective orbital and spin motions lead to a small change in the magnetic moment of the metal and induces off-diagonal terms in the optical conductivity tensor. The IFE is a phenomena of fundamental importance to the field of ultrafast magnetism [3,4]. Additionally, because IFE can enable light to exert torques on a magnetic materials, it has potential applications in opto-magnetic devices [5,6]. As a result, the IFE has been the subject of a number of recent experimental [7–10] and theoretical studies [2,11–14]. However, gaps in knowledge remain. For example, while theoretical studies have made predictions about how the strength of the effect varies from metal to metal, there are only a handful of experimental tests of this subject [6–8]. There have not been experimental studies of the IFE in metals like Ta and W, which are both known to exhibit "giant" spin-orbit transport properties in response to DC electric fields [15,16].

In this paper, we report time resolved measurements of the real and imaginary Kerr angle of a metal in response to excitation with circularly polarized (CP) light. (Following the convention of prior studies of this phenomenon [8,13], we use the term "specular inverse Faraday effect" to describe the Kerr response induced by CP pump light.) The goal of our study is to learn how different material properties of metals govern the magnitude of the specular IFE. We perform experiments on Cu, Au, Pd, Pt, α-W, β-W, α-Ta, and β-Ta thin-films. These metals were selected because of their differences in optical properties, spin-orbit coupling, and crystal structure. We observe that, despite significant differences in crystal structure, optical properties, electrical conductivity, and spin-orbit interactions, the IFE is similar in most of the metals we study. Excitation of Cu, Au, Pd, and Pt with a circularly polarized laser pulse with sub-ps duration induces a Kerr angle in these metals between 1.2 and 1.6 $\frac{nrad}{GW\ m^{-2}}$. The response of β-Ta, α-Ta and β-W is slightly larger. In these metals, we observe a Kerr response of ≈ 2, 2, and 5 $\frac{nrad}{GW\ m^{-2}}$, respectively. Alternatively, α-W has a Kerr response that is an order of magnitude larger than other metals. For α-W, we observe a Kerr response of ≈ 20 $\frac{nrad}{GW\ m^{-2}}$. We credit the large specular IFE to large spin-orbit coupling of 0.5 eV for W bands [17], together with an interband transition energy near our laser energy of 1.58 eV [17,18].

## Methods

Sample Preparation

We prepared metallic thin films on (0001)-oriented double-side polished $Al_2O_3$ substrates. We cleaned the substrates thoroughly with acetone and isopropyl alcohol, and then we annealed the substrates at 1100 °C in air for an hour to improve the surface quality. Right after annealing, we deposited 20 nm of the material (Au, Cu, Pt, Pd, W, Ta) using an AJA Orion sputtering system, in a 3.5 mTorr Ar atmosphere. During deposition, the sample stage was rotated during deposition at 20 rpm. All metals were grown at room temperature except for the α phases of Ta and W, which were grown at 850°C and then cooled to room temperature at a rate of ≈ 5 °C/minute. We then capped all samples (except for Au and Pt) with a 3 nm amorphous $Al_2O_3$ layer.

Characterization

To assess crystal structure of the samples, we performed X-ray diffraction measurements at room temperature using a PAN-alytical Empyrean diffractometer with Cu $K_\alpha$ radiation and a Ni filter, in the 2θ range 30°-110°. For all samples, we observed the peaks corresponding to the (0006) and (00012) Bragg peaks for the $Al_2O_3$ substrate, along with the corresponding peaks for the deposited materials (**Figure 1a**). For the cases of Cu, Pd, Pt and Au, we observed the characteristic face-centered cubic (FCC) crystalline phase with the diffraction peak corresponding to the (111) orientation. For the cases of Ta and W grown at 850°C, the (011) and (022) peaks corresponding the body centered cubic (BCC) α-phase were observed. Lastly, for the cases where Ta and W grown at room temperature, peaks corresponding to reflections from (002), (022), (004) and (044) lattice planes, corresponding to the tetragonal β-phase are observed. The broadening of the XRD peaks hints the formation of a strained polycrystalline array. We also verified the morphology of the thin films surface by atomic force microscopy (AFM), we found that the surfaces were flat and uniform, with root-mean-square (RMS) roughness less than 1 nm RMS (**Figure 1b**). Finally, we measured the resistivity of the thin films by four-point probe method. These values are reported in **Table I**.

Pump Probe Measurements

For the time-domain thermoreflectance (TDTR) and time-resolved magneto-optic Kerr effect (TR-MOKE) measurements, we used a pump-probe system built around a Ti:sapphire laser with a repetition rate of 80 MHz [19]. An electro-optic modulator (EOM) modulates the pump laser beam at a frequency of $f_{mod}$ = 10.7 MHz. The specular IFE signal is expected to have a temporal shape proportional to the convolution of the pump and probe beam intensities. The convolution of the pump and probe pulses in our apparatus has a $\text{sech}^2\left(\frac{t}{\tau}\right)$ time shape, with τ = 330 fs, FWHM = 600 fs for λ = 783 nm. A delay stage varies the arrival time of the pump laser pulses to the sample relative to the probe pulses (**Figure 2**). The pump beam fluence in these experiments is set to 5.1 J/m$^2$. The 1/e$^2$ radius of the focused pump and probe beams is 5.6 μm. The sample is mounted in such a way that the pump beam goes through the substrate first and impinges on the metallic film at metal/sapphire interface, i.e., the bottom of the metal film. The probe beam is incident on the top surface of the metal film. The angle of incidence with the sample is ≈ 0° for both pump and probe beams. The pump beam polarization is set to be right circularly polarized (RCP) or left circularly polarized (LCP) with a super achromatic quarter-wave plate. The probe beam polarization is linearly polarized in the horizontal plane. This pump/probe polarization configuration makes contributions to our signal from the ordinary Kerr effect negligible [7,8]. To independently measure the real (rotation) and imaginary (ellipticity) components of the IFE induced Kerr angle in our TR-MOKE measurements, we utilized a liquid crystal variable retarder (LCVR) which was adjusted to provide the needed retardance to separate both components and to account for the added ellipticity from the reflecting surfaces in the detection line [20]. The probe beam reflected from the sample is split into orthogonal polarizations and focused onto one of two photodiodes in a balanced photodetector. Balanced photodetector voltage is measured with an RF lock-in amplifier as a function of time-delay. We conduct a pump/probe scan with an LCP pump beam and then a separate scan with an RCP pump beam. Both scans typically include a small background signal that is unrelated to IFE due to imperfect balancing of the balanced detector. Finally, to obtain the total IFE signal, we subtract the signals collected with LCP and RCP pump beams. We note that, while the center wavelength of our laser is 1.58 eV, or 783 nm, our pump/probe measurements use a two-tint system to prevent pump light from reaching the photodetector. In this two-tint system, we use sharp edged high- and low-pass optical filters to slightly red- and blue-shift the spectrum of the pump and probe beams by ≈ 3nm, respectively.

The time-scale for magnetization to build up or dissipate in response to CP light being turned on or off is < 10 fs [7,21]. Since the sub-picosecond pulse duration in our experiments is much longer than this, we are measuring the specular inverse Faraday effect under quasi-steady-state conditions.

## Results

The metals we studied fall into one of three groups. The first group are metals in the periodic table group 10, with nearly filled d-electron bands (Pd, Pt). The second group are metals in group 11, with electrons in s-bands (Cu, Au). The third group are metals in groups 5 and 6, with electronic d-shells partially filled (Ta, W).

The measured Kerr angles (the real part being the Kerr rotation $\theta_K$, and the imaginary part the ellipticity $\varepsilon_K$) are shown in **Figure 3** and **Figure 4**. A positive $\theta_K$ indicates a counterclockwise rotation of the major axis of polarization (in the probe beam system of reference), while a negative $\theta_K$ corresponds to a clockwise rotation. Similarly, a positive $\varepsilon_K$ represents an increase in the right circular polarization of the beam, while a negative $\varepsilon_K$ indicates an increase in the left circular polarization. The results shown in Figs. 3-5 show the Kerr rotation in response to a constant incident fluence of 5.1 J/m$^2$.

The metals in groups 10 and 11 all have an FCC lattice structure. Cu and Au have filled d-bands and are "free-electron-like" metals. Pd and Pt have nearly filled d-bands. For all these samples, the Kerr rotation is positive. Alternatively, for the ellipticity part, we observed at 1.58 eV that Cu, Pd and Pt have a measurable positive signal, while Au displayed a very small negative signal.

The metals in groups 5 and 6 have partially filled d-bands. As mentioned in the characterization section, depending on deposition temperature, sputter deposition of W and Ta thin films yields either a body-centered-cubic α-phase or a non-cubic β-phase when the process is performed at low temperatures. The non-cubic β-phase typically has less crystal symmetry (more disorder) and much lower electrical conductivity. This allows us to test the effect of high vs. low electrical conductivity on the IFE signals by comparing measurements of the β-phase films to the BCC α-phase. Compared to the FCC metals previously mentioned, the β-Ta sample had a negative value for the $θ_K$, while $ε_K$ had a positive value. The values are similar in magnitude but opposite sign to the ones observed in Pt. For α-Ta, the Kerr angle values are approximately twice those observed in the β-Ta. Alternatively, the β-W sample showed Kerr angles similar in magnitude (~5μrad) compared to all the previously studied FCC metals, while the α-W sample the resulting Kerr angles are one order of magnitude larger than the β-W sample.

As noted in methods, the circularly polarized pump beam goes through the sapphire substrate to excite the metallic film. To confirm our measured signals result from the metal film, and not the substrate, we performed the IFE experiment on an Au film grown on MgO (001) that was sputter deposited simultaneously as the Au/sapphire sample whose IFE signals are shown in **Figure 3**. If our signals shown in **Figure 3** or **4** contained non-negligible contributions related to the nonlinear optical properties of the substrate, then the measured IFE signal should change if the substrate is changed. However, we observed nearly identical Kerr angles ($\theta_K$, $\varepsilon_K$) for Au on MgO and Au on sapphire [22]. Interestingly, while the signal magnitude was similar, the signal to noise was better for the Au/MgO signal, which may be related to order at the film/substrate interface.

## Discussion

The temporal shape we observe for the specular IFE signal of all metals is similar. The recorded data was fitted to account for artifacts in the signal [23], and we use the obtained amplitudes for the comparison among all the performed measurements.

For the four FCC metals (Cu, Au, Pt, Pd), we observe a similar magnitude of the specular IFE. Being isovalent metals, Au and Cu have similar band-structure, as do Pt and Pd. However, due to their larger atomic number, Pt and Au possess larger spin-orbit coupling than Pd and Cu. Our experimental results indicate there is a slightly larger specular IFE in Au than Cu, and in Pt than Pd (**Figure 3**). Nonetheless, we observe that the effect of a change in elemental mass, and hence spin-orbit coupling strength, is not necessarily large.

The groups of metals we studied have systematic differences in band-structure. Group 5 and 6 elements (Ta and W) are transition metals with mostly empty d-states. Group 10 elements (Pd and Pt) are transition metals with mostly full d-states. Group 11 elements (Cu and Au) are noble metals with filled d-states. As a result, these 6 metals have distinct electrical and optical properties. The electrical resistivities of our Pd, Pt, W and Ta thin films are larger compared to Cu and Au (**Table I**). Furthermore, at the laser wavelength of 786 nm, the absorption of Cu and Au is much lower than the other studied metals. Our initial hypothesis when beginning our experiments is that these differences in electronic band-structure, electrical properties, and optical properties would lead to significant differences in the specular IFE. However, we did not observe any clear correlation between absorptance or electrical conductivity and the measured specular IFE signals. To compare how the specular IFE response varies between the metals we studied, we normalize the measured Kerr angles to the total incident intensity of the CP pump beam, and plot the real and imaginary Kerr response in a polar plot in **Figure 5**, revealing that the IFE signal is similar in all metals we studied, with the exception of W and Ta.

The group 5 and 6 metals (Ta and W) demonstrated a larger specular IFE signal magnitude than the group 10 and 11 FCC metals examined in this work. By exploring the metastable β-phases of Ta and W, we were able to examine the impact of crystalline order and electrical conductivity on the specular IFE. Previous studies have indicated that thin-films of β-phase Ta and W exhibit lower crystalline symmetry in comparison to the α-phase Ta and W films, which are deposited at higher temperatures [24–26]. Consequently, β-phase Ta and W demonstrate higher electrical resistivities and reduced optical reflectance compared to their corresponding α-phases, as noted in **Table I**. The differences in phase, order, and electrical conductivity for these metals should produce a larger scattering rate in the oscillating electrons, hence reducing the magnitude of the IFE signal. Specifically, the α-phase Ta and W exhibit a Kerr response to circularly polarized pump light that is two and ten times larger than β-phase Ta and W, respectively. We note that, in addition to the differences in scattering rates, the α- vs. β -phases of Ta and W have differences in band-structure that may also explain the observed differences in the specular IFE signals.

In order to understand the observed behavior among the studied thin film metals, we consider theoretical predictions for the magnetization induced by circularly polarized (CP) light. Notably, these theoretical models do not attempt to predict the Kerr effect resulting from the induced magnetic moment. The link between magnetization and Kerr rotation is multifaceted, influenced by factors like the electronic states that are excited by CP light, and their respective orbital and spin moments. Hence, a direct theory-experiment comparison is beyond the scope of the current work. We focus instead on identifying any similar trends, i.e., whether theory predicts a larger induced magnetic moment in metals we observe a large IFE signal.

According to the semiclassical theory outlined by Hertel, the magnitude of the induced magnetization caused by orbital motion of free electrons is proportional to the plasma frequency $\omega_p$ of the metal [2,27]. We plotted the expected magnetization per irradiance unit $\left(\frac{dM}{dI}\right)$ vs. the studied material, and we compared this with our results for the measured Kerr rotation per irradiance $\left(\frac{d\theta_k}{dI}\right)$ (**Figure 6**), in order to determine if there is a trend between these parameters. We can first observe that the semiclassical approach does not reproduce the change of signs in the observed Kerr rotations, hence a more comprehensive approach is required.

Hertel's semi-classical theory of IFE on metals only includes orbital contributions to the IFE made by free-electrons. In addition to these intra-band orbital moments, circularly polarized light will induce spin and orbital moments due to interband transitions. To evaluate whether differences in the spin-moments induced by interband transitions could explain our experimental trends, we performed density functional theory (DFT) calculations following the procedure outlined in Ref. [28]. The agreement between density functional theory calculations and experiment is mixed. Consistent with our experimental observations of the specular IFE, density functional theory predictions of the spin-moment are comparable for Cu, Pd, α-Ta, and Au. However, the DFT calculations predict that Pt should have a large spin-moment induced by CP light, while we observe that Pt has a specular IFE comparable to Cu, Pd, and Au.

There are multiple possible explanations for the discrepancy between Hertel and DFT theories and our experimental observations. The magnetization induced by IFE in the studied metals may have a sizeable contribution from both orbital and intraband electronic contributions. Another possibility is that the proportionality constant to convert the Kerr rotation $\theta_k$ to induced magnetization M might be different due to optical resonance effects, as it has been observed in thermoreflectance experiments for tungsten [17]. In prior work, we observed that the Kerr rotation caused by spin accumulation in Au was a strong function of wavelength due to interband transition resonances [20]. We expect the same to be to be true for spin and orbital moments induced by circularly polarized light.

We now compare our results to prior studies of the specular IFE. Similar to our study, several prior experimental studies report that many metals display similar specular IFE signals. Kruglyak *et al.* report the specular IFE for thin-films of Au, Cu, Ag, Pd, Ni, Hf, Zr, and Ti deposited on silicon [8]. We do not expect identical results to Kruglyak et al. because a) their experiments were not conducted at normal incidence, and b) they used pulses with a 0.1 ps duration, which means their experiments are not quasi-steady-state like ours. Kruglyak observed a specular IFE magnitude of 0.05 $\frac{nrad}{GW\,m^{-2}}$ for Cu, 0.37 $\frac{nrad}{GW\,m^{-2}}$ for Au and 0.62 $\frac{nrad}{GW\,m^{-2}}$ for Pd. angles of incident. In the work of Kruglyak et al., Zr displayed largest specular IFE magnitude (1.26 $\frac{nrad}{GW\,m^{-2}}$), which is two times the value they observed for Pd. The fact that metals with partially filled d-levels have larger specular IFE than metals with fully filled d-level is in agreement with our experiments. The specular IFE in Zr is comparable to α-Ta, β-W, β-Ta. The signals we observe for α-W are 15 times higher than Zr after accounting for differences in fluence and pulse duration in the two experiments.

The magnitude of the specular IFE that we observe for W is greater than other prior reports of large specular IFE signals. In the case of plasmonic gold nanoparticles [29], the maximum observed value was 3 $\frac{nrad}{GW\,m^{-2}}$ at a resonance energy of ~ 2.18 eV. We note that this specular IFE was observed by transmitting the laser pulse through 0.2 cm of nanoparticle solution, while the probe beam in our experiment is passing through only 20 nm of metal. Kozhaev *et al.* explored using optical resonances engineered via nanocavities to enhance IFE. They observed 1.4 $\frac{nrad}{GW\,m^{-2}}$ when a magnetophonic microcavities fabricated from Bi-doped iron garnets was excited with CP light [30]. Another interesting comparison for our result for W is to $Tb_3Ga_5O_{12}$. $Tb_3Ga_5O_{12}$ is a material known for its strong magneto-optical properties. Mikhaylovskiy et al. reported an IFE signal of ≈ 9 $\frac{nrad}{GW\,m^{-2}}$ for a 1mm thick single crystal excited with CP light at 1.55 eV [31]. The large IFE signals we observe suggest α-W could be an exceptionally good transducer for magneto-optical devices that are designed to optimize the torque circularly polarized light exerts on a magnetic heterostructure [6,9].

In conclusion, we measured the specular IFE signal at 1.58 eV for different transition metals thin films. We found that the materials with partially occupied d-shell electron levels exhibit the larger specular IFE signals, and that films with materials with more ordered phases have stronger responses compared to those with disordered phases. The magnitude of the measured IFE signals does not appear to follow the trend predicted by the semiclassical IFE model; however, a more comprehensive theoretical model is needed to properly correlate the expected Kerr rotation signal to the irradiated metal. The extraordinarily large specular IFE response in the α-W film, which exceeds any of the other materials studied previously, might be caused by the angular momentum of the electrons at the 1.58 eV energy band, thus photon-energy dependent measurements are required to understand the nature of this behavior. Finally, the results reported in this work represent an opportunity to deepen our understanding in photoinduced magnetization in thin film materials, which could be translated to the design and implementation of novel optospintronic devices.

## Acknowledgements

The experimental work was supported by the U.S. Army Research Laboratory and the U.S. Army Research Office under Contracts/Grants No W911NF-20-1-0247. The theoretical work was also supported by NSF Grant No. DMR-1848074. The authors declare no conflict of interest.

| Material | Roughness (Å RMS) | ρ (nΩ m) | Absorptance | θ$_K$ μrad | ε$_K$ μrad | $\frac{d\Phi_K}{dI}$ $\frac{nrad}{GW\ m^{-2}}$ | $\hbar\omega_p$ (eV) |
|---|---|---|---|---|---|---|---|
| Cu | 9.9 | 13 | 0.05 | 5.4 | 2.0 | 1.4 | 7.34 |
| Au | 3.2 | 15 | 0.06 | 6.4 | -0.4 | 1.6 | 9.03 |
| Pd | 1.3 | 55 | 0.28 | 4.6 | 1.2 | 1.2 | 5.46 |
| Pt | 3.0 | 50 | 0.28 | 5.4 | 2.6 | 1.5 | 5.15 |
| β-Ta | 1.7 | 90 | 0.35 | -6.6 | 5.9 | 2.2 | |
| α-Ta | 6.3 | 70 | 0.36 | -16.5 | 9.2 | 4.7 | 8.6 |
| β-W | 2.3 | 420 | 0.34 | 4.5 | -8.9 | 2.5 | |
| α-W | 1.2 | 46 | 0.36 | 57.0 | -56 | 20 | 6.41 |

**Table I**: *Optical properties of the materials studied in this work. The plasma frequency $\hbar\omega_p$ was obtained from Ref. [32] and Ref. [33]*

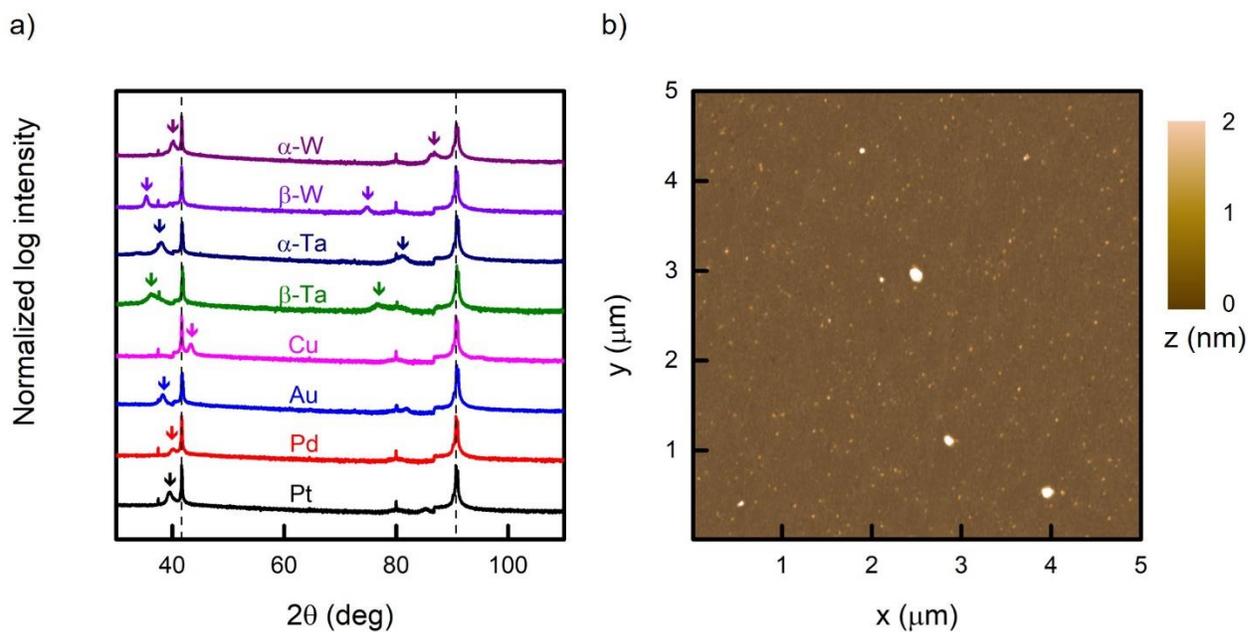

*Figure 1 Structural and surface morphology characterization of thin film samples.* a) X-ray diffractograms of the different metallic thin films grown on Al$_2$O$_3$ (0001)-oriented substrate. The dashed lines indicate the diffraction peaks corresponding to the (0 0 0 6) and (0 0 0 12) plane reflections of the substrate. The arrows indicate the peaks corresponding to the metallic thin films, indicating the samples are texturized in the (111) orientation. b) Atomic force microscopy image of the 20 nm β-W sample, with a root-mean square roughness of 2.3 Å.

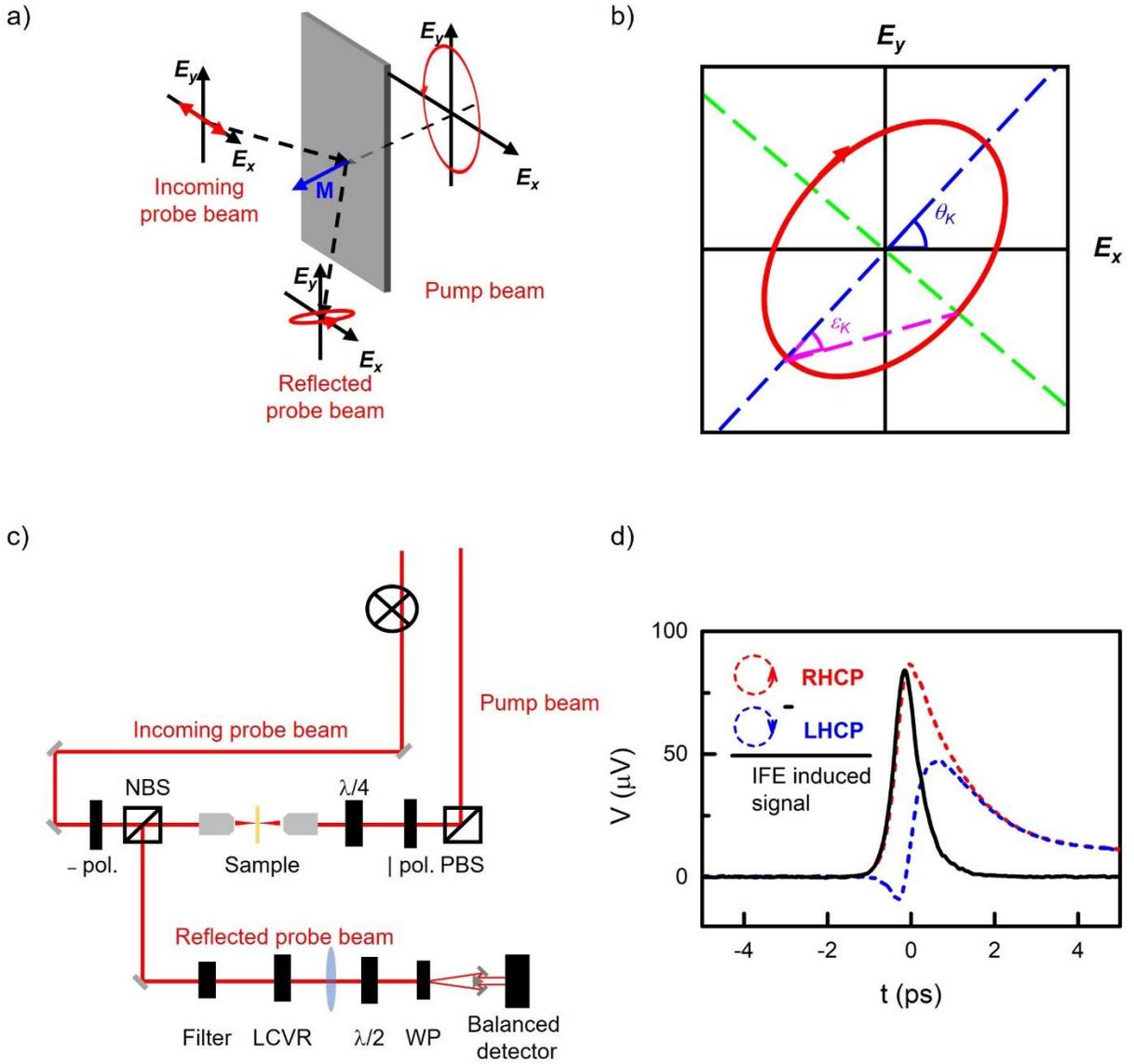

***Figure 2 Inverse Faraday Effect experiment.*** *a) Description of the Inverse Faraday Effect experiment: a thin film of the material is excited by a circular polarized pump laser beam pulse, which will induce a magnetization perpendicular to the film plane. A linearly polarized probe beam pulse will reflect, and it will change its polarization status. b) Diagram describing the Kerr rotation ($\theta_K$) and ellipticity ($\varepsilon_K$) angles measured in this experiment. c) Schematic of the pump/probe experimental set up. The pump beam is polarized circularly by means of a super achromatic quarter wave plate ($\lambda/4$), the linearly polarized probe beam reflects back into the detection line. d) Time-resolved measured signals by the balanced detector when using right-hand circular polarized and left-hand circular polarized pump beams to excite the sample. By subtracting both signals the total contribution of the IFE is then obtained.*

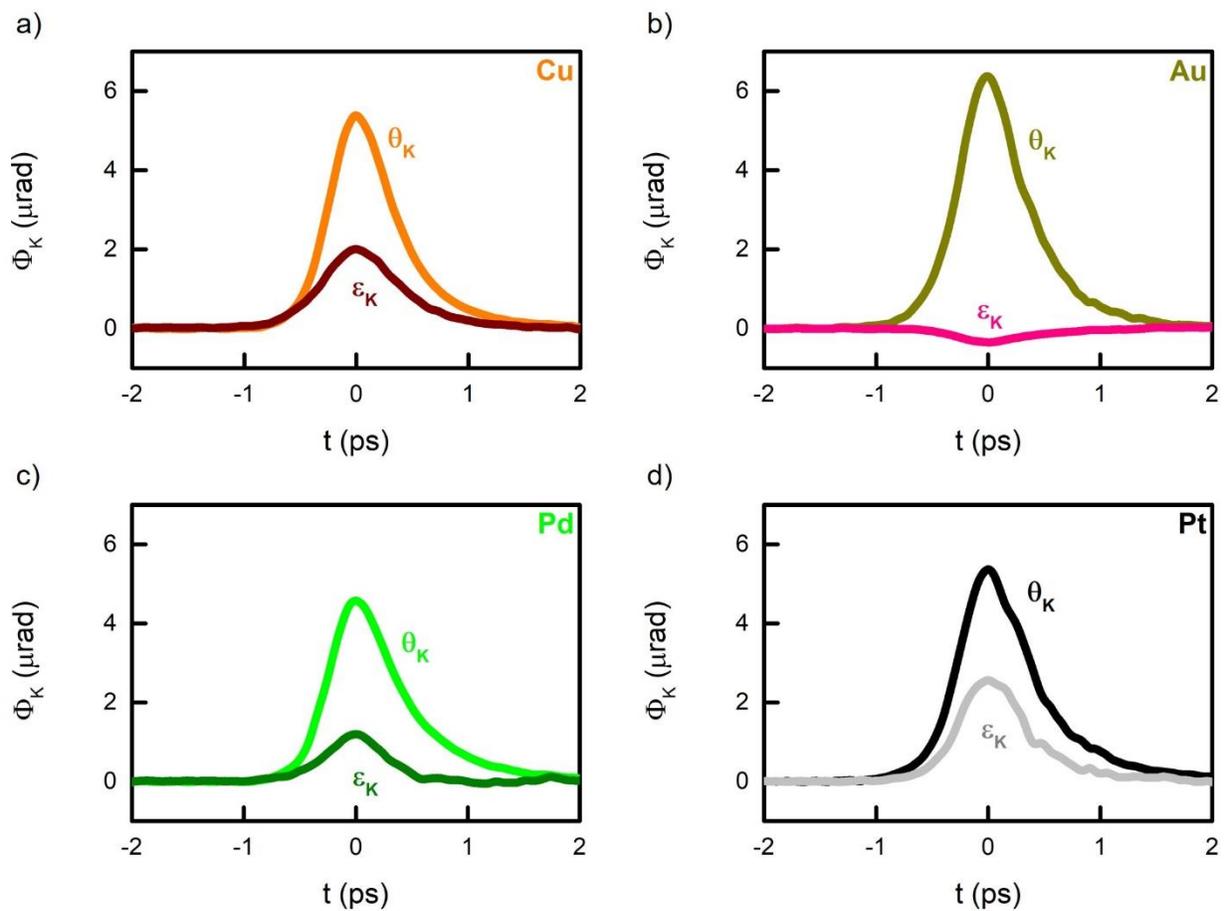

*Figure 3 Inverse Faraday Effect on VIIIB and IB metals.* Kerr rotation ($\theta_K$) and ellipticity ($\varepsilon_K$) for 1.58 eV photon energy for different face-centered cubic transition metals.

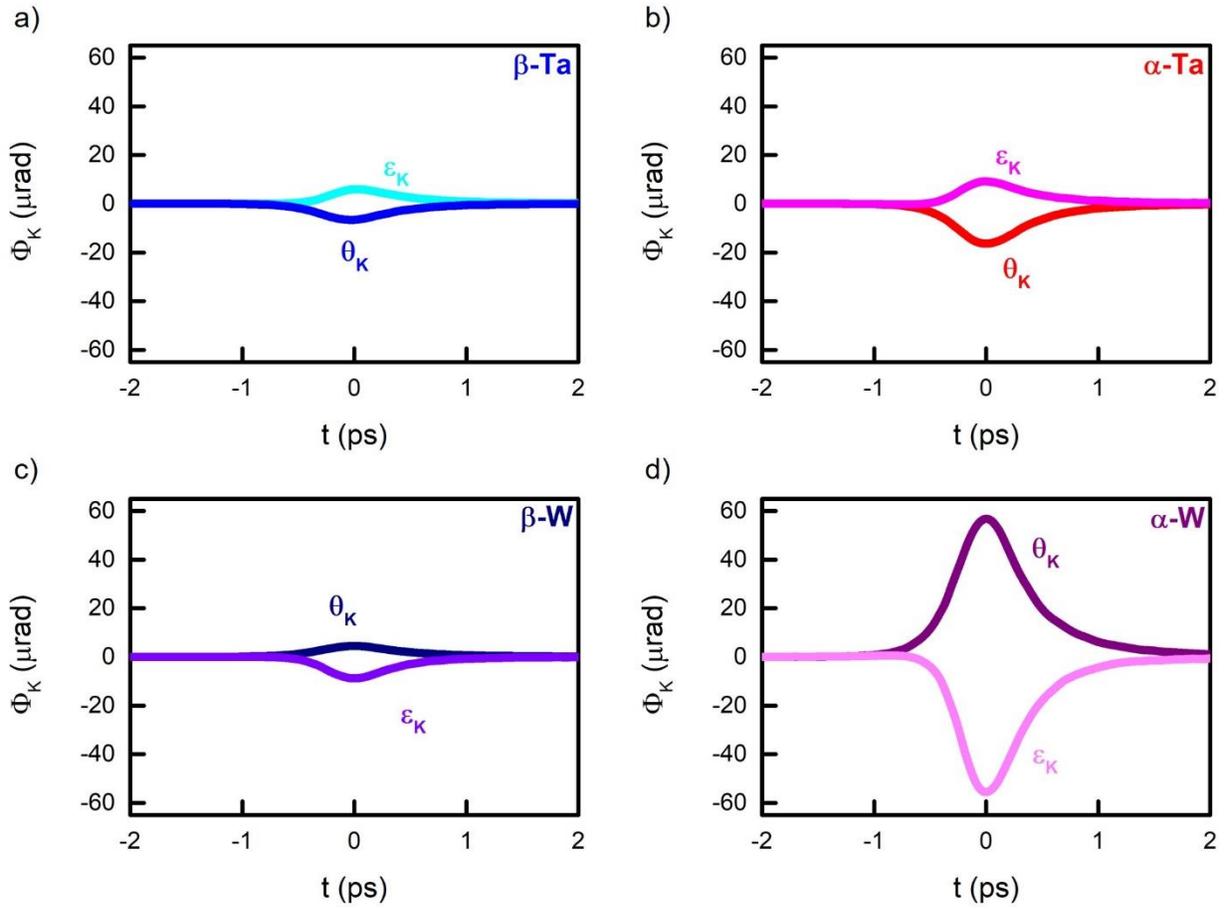

***Figure 4 Inverse Faraday effect on VB-VIB metals.*** *Kerr rotation ($\theta_K$) and ellipticity ($\varepsilon_K$) 1.58 eV photon energy for a) & b) tantalum and c) & d) tungsten.*

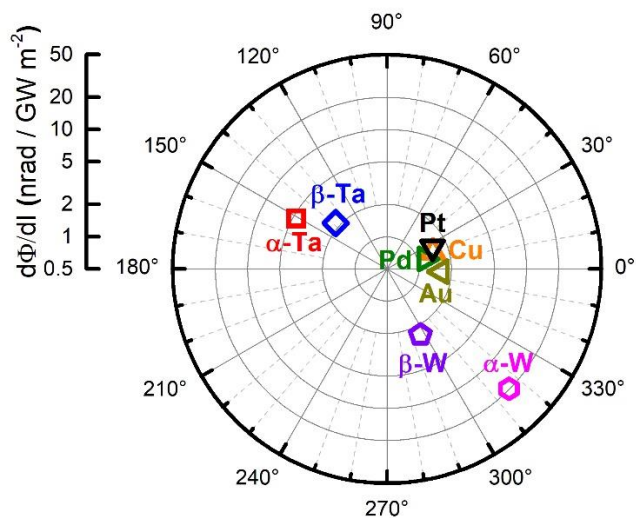

*Figure 5 Kerr angles per irradiance induced by the inverse Faraday effect.* *These values have been normalized by the exciting pump beam parameters, in order to express the measured values as an extrinsic material property. The horizontal axis corresponds to the Kerr rotation (real part), while the vertical axis corresponds to the ellipticity (imaginary part).*

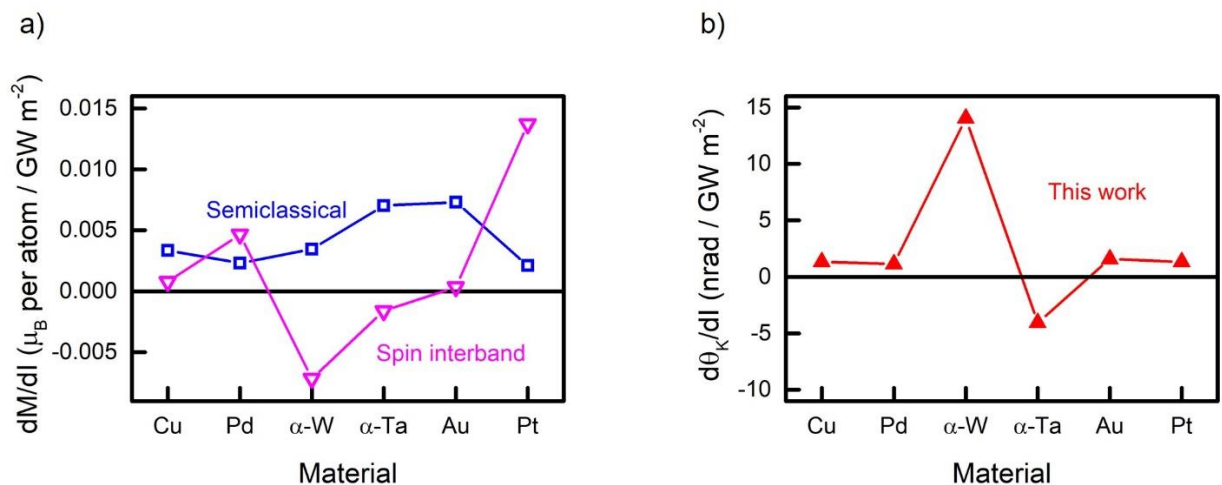

*Figure 6 a) Theoretical models and b) experimental data trends. Comparison between expected values for magnetization per irradiance according to theoretical models (semiclassical theory for IFE* [2] *and electronic spin contributions* [28]*), and our experimental values.*

(2006).